\documentclass[twocolumn]{article} 
\usepackage{graphicx}
\usepackage{txfonts}
\usepackage{natbib}
\usepackage{float}
\begin{document}
\title{Search for a correlation between ANTARES neutrinos and
       Pierre Auger Observatory UHECRs arrival directions}

\author{The ANTARES collaboration}


\date{28.02.2012}

\maketitle
 

{\bf This paper presents a search for correlation in the arrival directions of
2190 neutrino candidate events detected in 2007-2008 by the ANTARES telescope, 
and 69 ultra-high energy cosmic rays (UHECRs) observed by the Pierre Auger Observatory between 
January 1st 2004 and December 31st 2009. No significant correlation was found.
The corresponding 90\% C.L. upper limit on the neutrino flux from each observed UHECR direction (assuming an equal flux from all of them
and for $E^{-2}$ energy spectrum) is 4.99$\times$10${^{-8}}$GeVcm$^{-2}$s$^{-1}$.}

\section{\large Introduction}

The astrophysical sources of ultra-high energy cosmic rays (UHECRs) remain unknown.
These particles are expected to be of extragalactic origin and potential sites of acceleration include
jets of gamma-ray bursts \citep{1995PhRvL..75..386W,1995ApJ...453..883V,1997PhRvL..78.2292W,2006ApJ...651L...5M} or
active galactic nuclei \citep{1987ApJ...322..643B,1991PhRvL..66.2697S, 1993A&A...272..161R,1993PhRvD..47.5270N}.
The search for UHECR sources is complicated by their
deflection in magnetic fields inside and outside of the Galaxy and by the
extremely low rates of these particles: above 10$^{19}$eV, about 1 particle per km$^2$ per century.
Due to their interactions with photons of the cosmic microwave background via the GZK mechanism \citep{greisen,zatsepin}, 
UHECRs propagation distances are limited, for example for protons, to about 100 Mpc.
While the existence of a cut-off in the energy spectrum of UHECRs, first observed by the
HiRes experiment \citep{2008PhRvL.100j1101A,2009APh....32...53T}, has now been
confirmed by the data of the Pierre Auger Observatory \citep{2008PhRvL.101f1101A,2010PhLB..685..239A},
the composition of the cosmic rays above a few 10$^{18}$eV, crucial for estimation of
expected magnetic deflection magnitude, remains
uncertain. Although data from the Pierre Auger Observatory seem to indicate a
transition from light to heavier composition above 40EeV, this trend is still subject
to large uncertainties, in particular related to the lack of accurate modeling of
hadronic interactions in the relevant energy domain. 

A new and promising method to study the origin of the UHECRs is the multimessenger approach, 
which is based on the detection of secondary fluxes of gamma-rays and
neutrinos associated with the
decay of pions resulting from the interaction of the UHECRs with matter or photon 
fields in the vicinity of the cosmic accelerators 
\citep{1999PhRvD..59b3002W,2001PhRvD..64b3002B,2008PhR...458..173B,2009APh....31..138B}.
Although gamma-rays have been linked to astrophysical sources by recent observations
(H.E.S.S., MAGIC, VERITAS, Fermi), an unambiguous
identification of these sources as sites of hadronic acceleration requires the detection of the
associated neutrino flux. Neutrinos, being neutral and  weakly interacting particles, are
neither deflected nor attenuated during their propagation from the source to the Earth. 
Their small cross section interaction with matter makes
their detection challenging and requires the construction of very large telescopes. 
The currently operating neutrino telescopes, ANTARES, IceCube and BAIKAL, 
have not yet observed any statistically significant cosmic neutrino source
\citep{2011ApJ...743L..14A,2011ApJ...732...18A,2009AstL...35..651A}.

In this paper, the first source stacking method optimized for a correlation of arrival directions 
of UHECRs and neutrinos have been developed and applied
on the neutrino candidate events detected
by the ANTARES telescope and the
UHECR events observed by the Pierre Auger Observatory.
Would such a correlation be observed, it would
indicate regions of the sky where the sources of UHECR and/or
neutrinos could plausibly lie, as well as shed light
both on the UHECR composition and on the intensity of magnetic
fields in and outside of the Galaxy.
An observed correlation would also exclude the possibility
that the dominant sources of UHECRs are transient sources, since the time delay between neutrinos and
protons coming from such a source, is orders of magnitude larger than the observation time
of the ANTARES telescope and the Pierre Auger Observatory.

This paper is organized as follows. The discussion about deflection of UHECRs in magnetic fields is presented in Section 2.
The data samples are presented in Section 3
and the background and signal simulations are explained in Section 4.
The angular search bin optimization and the discovery potential are discussed in Section 5
and the results are given in Section 6.

\section{\large Magnetic deflection of UHECRs}

In this paper, ANTARES telescope neutrino candidate events directions were correlated with UHECR events
recorded by the Pierre Auger Observatory, using a source stacking method
in which the cumulative neutrino signal from
all UHECR directions is summed and compared with the expected background.
Such an approach is known to be sensitive to a significantly lower
flux per source for a 5 $\sigma$ 50\% discovery potential than the single point source
approach. A key parameter for the analysis is the angular search cone around each UHECR direction 
and is mainly determined by the assumed magnetic deflection of the 
UHECRs.   

Protons with the highest energies (above 10$^{19}$eV) are expected to be deflected by the Galactic magnetic field up to a few degrees \citep{1997ApJ...479..290S,
2002ApJ...572..185A,2010ApJ...724.1456T}.
\citet{1998ApJ...492..200M} calculated that protons with energies of 4${\times}$10$^{19}$eV
should be deflected by about 5${^{\circ}}$. \citet{1999JHEP...08..022H} concluded that 10$^{20}$eV protons arrive to Earth almost undeflected.
Deflection angles of about 3${^{\circ}}$, for protons of 4${\times}$10$^{19}$eV,
were estimated by \citet{2003A&A...410....1P}.
Also, most authors suggested insignificant deflection angles for protons traveling through extragalactic magnetic fields,
even for a propagation through galaxy clusters \citep{2005JCAP...01..009D,2008PhRvD..77l3003K,2008JPhCS.120f2025D,2010PhRvD..82d3002A}.
However, if the composition of UHECRs is mostly heavy, identification of their sources would be likely impossible.
\citet{1998ApJ...492..200M} found that Fe nuclei with energy of
2.5${\times}$10$^{20}$eV can be deflected up to 20${^{\circ}}$ in the Galactic magnetic field. \citet{2003A&A...410....1P} also calculated
deflection angles of few tens of degrees for heavy UHECRs. This was also confirmed in a recent paper by \citet{2010ApJ...724.1456T}.

In order to avoid the trial factor associated with using multiple tolerance windows for the 
magnetic deflection a single value of 3$^{\circ}$ is adopted for this analysis.

\section{\large Neutrino and UHECR data}

The ANTARES neutrino telescope \citep{2011NIMPA.656...11A} is located in
the Mediterranean Sea, 40 km off the southern coast of France (42$^{\circ}$ 48$^{'}$N, 6$^{\circ}$ 10$^{'}$E), at a depth of 2475 m.
It was completed in 2008 and its final configuration is 
a three-dimensional array of photomultipliers in glass spheres (optical modules \citep{2002NIMPA.484..369T}),
distributed along twelve lines anchored at the sea bottom and kept taut by a buoy at the top. 
Eleven of these detection lines contain 25 storeys of triplets of 
optical modules and one contains 20 triplets. The lines are subject to the sea currents and can 
change shape and orientation. A positioning system based on hydrophones, 
compasses and tiltmeters is used to monitor the detector geometry with an accuracy of 10 cm. 
The total instrumented volume of the ANTARES telescope is about 10${^7}$m$^3$. 
The detection principle is based on measuring the Cherenkov light emitted in the detector by high energy muons, that result from neutrino
interactions inside or near the instrumented volume of the detector.
The large background from downgoing muons produced in cosmic ray
air showers is reduced by selecting only upgoing muons as neutrino candidates.

The data acquisition system of the detector \citep{2007NIMPA.570..107A} is based on the 
"all-data-to-shore" concept, in which signals from the
photomultipliers above a given threshold are digitized and sent to shore for processing.
The absolute time is provided by GPS and the precise timing resolution for the recorded 
photo-multiplier tube signals, of order 1 ns, is required to maintain 
the angular resolution of the telescope. The arrival times of the hits are calibrated as 
described in \citep{2011APh....34..539A}. 
A L1 hit is defined either as a high-charge hit, or as hits separated by less than 20 ns 
in optical modules of the same storey. At least five L1 hits are required 
throughout the detector within a time window of 2.2 ${\mu}$s, with the relative photon arrival 
times being compatible with the light coming from a relativistic particle. 
Independently, events which have L1 hits on two sets of adjacent or next-to-adjacent floors 
are also selected. The physics events are stored on disk for offline 
reconstruction.

The data used in this analysis were collected between Jan 31st, 2007 and Dec 30th 2008. During this
time the construction of the detector was still in progress.
The detector consisted of 5 lines for most of 2007 and of 9, 10 and 12 lines during 2008.
For part of that period, the data acquisition was interrupted for the connection of
new lines, and in addition, some periods were excluded due to the high bioluminescence-induced
optical background. The resulting effective live time of the analysis is 304
days.

These events were reconstructed offline to determine the muon trajectory, using a multi-stage 
fitting procedure. The final stage of this procedure 
consists of a maximum likelihood fit of the measured photon arrival times. 
A quality parameter, indicated by ${\lambda}$, is determined based on the final value of the 
likelihood function. 
Selection cut on parameter $\lambda>$-5.4 has been optimised in order to have the
best point source sensitivity \citep{2011ApJ...743L..14A}.
The estimated angular error obtained from the muon track fit is required to be 
smaller than 1$^{\circ}$.
The final data sample consists of 2190 up-going neutrino candidate events. 
For this current analysis, no selection was done based on the energy reconstruction. 
The angular resolution was estimated to be 0.5${\pm}$0.1$^{\circ}$.  The simulations indicate that 
the selected sample contains 60\% atmospheric neutrinos, the rest are misreconstructed atmospheric muons.

Previously, the Pierre Auger Observatory reported an anisotropy in the arrival directions of UHECRs
\citep{2008APh....29..188P} and indicated a
correlation with Active Galactic Nuclei (AGN) from Veron-Cetty\&Veron (VCV) catalog \citep{2006A&A...455..773V}. 
After a scan of the relevant parameters, the prescription was made on a subsample of data and 
the correlation was found to be the most
significant for a sample of 27 events corresponding to cosmic ray energies higher
than 57EeV, falling within a bin of size 3.1$^{\circ}$ around the AGNs from the VCV catalog, located
at distances smaller than 75Mpc.
However, the HiRes Collaboration reported an absence of a
comparable correlation in observations in the Northern hemisphere \citep{2008APh....30..175T}.
Further, the suggested correlation of the Pierre Auger UHECRs with the nearby AGN sources
decreased in the following analysis
\citep{2010APh....34..314T} with 69 events at energies above 55EeV (10$^{19.74}$ eV), observed until 31 December 2009.
These 69 UHECR events were used in the correlation analysis presented in this paper.

\begin{figure}[H]
\begin{center}
\includegraphics*[width=\columnwidth,clip]{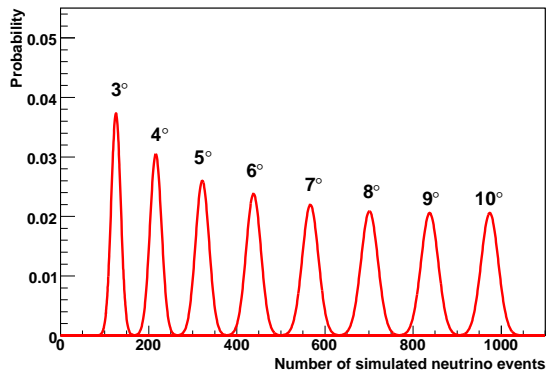}
\caption{\label{backg} The probability density functions of the number of neutrino events in
3-10$^{\circ}$ bins centered on 69 UHECR directions.}
\end{center}
\end{figure}

\section{\large Background and signal simulations}

In order to study the statistical significance of any observed correlation between datasets and
determine an optimal angular search bin, Monte Carlo (MC) set with 10$^{6}$ pseudo-experiments 
is generated, each with 2190 
neutrinos and 69 UHECRs. In each of these 
pseudo-experiments the positions of UHECRs are
fixed according to the Pierre Auger Observatory dataset and 
the neutrino background is randomly generated by scrambling the 2190 neutrinos
from the ANTARES telescope dataset in right ascension.
The number of neutrinos within an angular bin of chosen size, centered on 69 UHECR events is counted. The normalized probability density function 
(PDFs) is
calculated and fitted with a Gaussian distribution, to obtain the mean neutrino count and its
standard deviation expected from the 
randomized background samples.
This procedure is repeated for a range of different bin sizes. 

\begin{figure}[H]
\begin{center}
\includegraphics*[width=\columnwidth,clip]{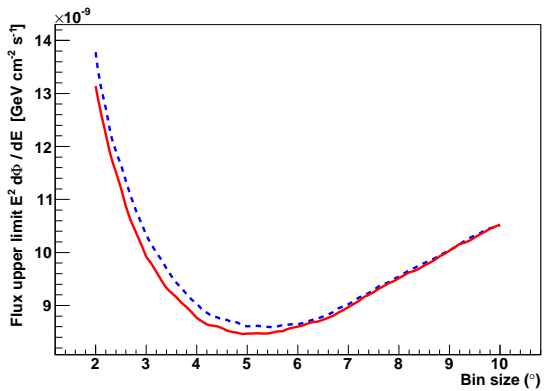}
\caption{\label{sens}The mean flux upper limit (90\% CL) as a function of the search bin
size is presented with the red solid line.
The mean flux upper limit for two times lower angular resolution is shown with the blue dashed line.
}
\end{center}
\end{figure}

For illustration,
Figure \ref{backg} shows an example count of neutrinos for bins of 3-10$^{\circ}$ size. The count of events is done 
by adding neutrinos in all 69 bins for which the minimum angular distance to UHECRs is smaller than the bin size. 
In this way, when the same neutrino event falls within multiple bins around the UHECRs, a double counting of neutrino events is avoided. 
After optimizing an angular bin size (as described in the next Section), the significance of the observed number of neutrino events within 69 bins 
is calculated by comparison with the distribution for the pure background MC sample.

The signal events are simulated assuming a neutrino energy spectrum proportional to 
$E^{-2}$ and equal flux strength from each of 69 UHECR directions. 
Flux values from 0.5$\times$10$^{-8}$GeVcm$^{-2}$s$^{-1}$ to 10$^{-7}$GeVcm$^{-2}$s$^{-1}$ are considered.
The flux is converted into signal event rate per source using the effective area for 5-12 lines and the corresponding live time.
For every source, signal neutrinos are generated according to the Poisson distribution with the event rate per
source as mean value. For example, a flux value of 10$^{-8}$GeVcm$^{-2}$s$^{-1}$, gives 0.85 signal neutrinos 
per UHECR source, or about 58 events
for all stacked sources.
Signal neutrinos are randomly generated 
according to a Gaussian which is a result of a convolution of the magnetic field
tolerance window of 3$^{\circ}$ and
the angular resolution of the ANTARES telescope.
The same amount of background neutrinos is removed from a declination band of 10${^{\circ}}$ centered on each UHECR to 
ensure that every random sky has 2190 events and to keep the neutrino declination distribution profile close to the observed profile.

\section{\large Angular search bin optimization and discovery potential}

Monte Carlo predictions are used to calculate the mean upper limit, or Feldman-Cousins sensitivity
\citep{1998PhRvD..57.3873F,2003APh....19..393H}, that would be observed over the set of pseudo-experiments 
with expected background $n_b$ and no true signal.
Over an ensemble of experiments with no true signal, the background $n_b$ will fluctuate to different values with different
Poisson probabilities, each one associated with an upper limit ${\mu}_{90}$. 
The mean upper limit is the sum of these expected upper limits, weighted by their Poisson probability of occurrence.
Over an ensemble of identical experiments, the strongest 
constraint on the expected signal flux corresponds to a set of cuts that
minimizes the model rejection factor (${\mu}_{90}/n_s$) and at the same time minimizes the mean flux upper
limit that would be obtained over the hypothetical experimental ensemble. The model rejection factor, as well as the optimized bin size, do not depend on
a chosen flux value, as both ${\mu}_{90}$ and $n_s$ are proportional to it.


The described Feldman-Cousin's approach with the Rolke extension \citep{2005NIMPA.551..493R}  
was used to calculate the mean upper limit on the neutrino flux per source assuming that the neutrino spectrum follows
$E^{-2}$, 
for a 90\% confidence level, from background samples as shown in Figure \ref{sens} (red solid line).
Using 3$^{\circ}$ magnetic deflection tolerance value, the angular search bin that minimizes the flux upper
limit is 4.9${^\circ}$.

\begin{figure}[H]
\begin{center}
\includegraphics*[width=\columnwidth,clip]{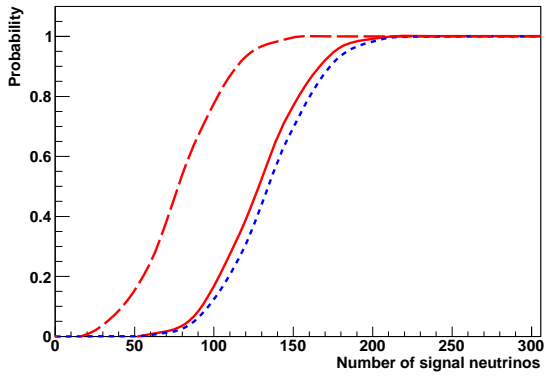}
\caption{\label{discpot1}The discovery potential at 3 ${\sigma}$ (red long-dashed line) and 5 ${\sigma}$ (red solid line)
90\% C.L. as a function of the number of neutrino signal events from 69 sources on the whole sky.
The discovery potential for two times lower angular resolution is shown with the blue dashed line.}
\end{center}
\end{figure}

With the angular search bin size optimized and fixed, it is possible to estimate the probability of making 
a 3 ${\sigma}$ or a 5 ${\sigma}$ 90\% C.L. discovery given a certain signal flux. 
First, the neutrino count necessary for a chosen $\sigma$ level is determined from the
background MC samples. 
Then, the number of pseudo-experiments with signal, that have more neutrinos in 69 
optimized bins than the chosen ${\sigma}$ level from background only, 
is counted and this gives a direct 
measure of the discovery potential for that particular flux. 
Figure \ref{discpot1} shows the discovery potential for 5${\sigma}$ (red solid line) and 3${\sigma}$ (red long-dashed line) discovery, for an optimized bin size 
of 4.9$^{\circ}$. 
Around 125 (75) signal events correlated to the 69 UHECRs directions are needed for a 5${\sigma}$ (3${\sigma}$) discovery in 50\% of trials.
This counts correspond to flux per source of 
2.16$\times$10$^{-8}$ GeV cm$^{-2}$ s$^{-1}$ and 1.29$\times$10$^{-8}$ GeV cm$^{-2}$ s$^{-1}$ 
respectively.

To check the effect of a possible angular resolution systematic error on sensitivity and discovery potential,
MC simulations with an angular resolution lowered by a factor of 2 were performed. The optimized bin value in this case is 5.5${^{\circ}}$, 
compared with 4.9${^{\circ}}$ obtained from the observed angular resolution. This 100\% larger angular resolution results in about 20\% higher
neutrino flux upper limit.
No significant effect is found on the discovery potential.
Figures \ref{sens} and \ref{discpot1} show respectively optimized bin 
and discovery potential for observed and 2 times lower angular resolution.
Note that the expected error on the angular resolution, as we already mentioned, is estimated to be much smaller ($0.1^{\circ}$).

\begin{figure}
\includegraphics*[width=\columnwidth,clip]{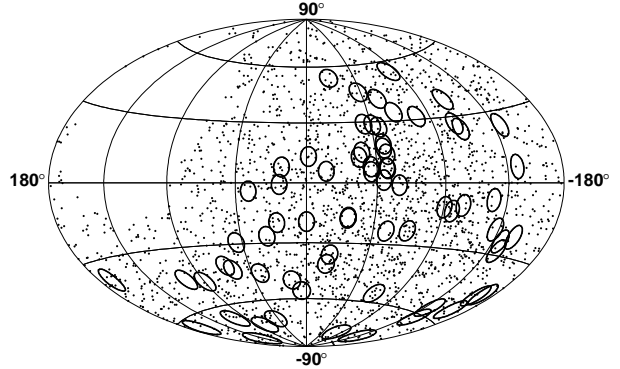}
\caption{\label{skymap}On this skymap in Galactic coordinates, neutrino events are represented with black dots
and angular search bins of 4.9${^{\circ}}$ centered on the
observed UHECRs with black circles.
}
\end{figure}

\section{\large Results}
To analyze the level of correlation between the distribution of 2190 neutrino candidates observed by ANTARES telescope,
and 69 UHECRs reported by the Pierre Auger Observatory, 
neutrino candidate events were unblinded.
The significance of an observed correlation is determined with the help of randomized background samples, using the optimized 
bin of 4.9$^{\circ}$. The most probable count for this optimized bin size, or the 
mean background expectation from the randomized samples, is 310.5 events (in all 69 bins), 
with the standard deviation of 15.2 events.
After unblinding 2190 ANTARES telescope neutrino candidate events, a count of 290 events 
within 69 bins is obtained (Figure \ref{skymap}), which is sligtly lower than expected.   
This count is  
compatible with a underfluctuation of the background, with a significance of 1.4 ${\sigma}$. 
The corresponding 90\% C.L. upper limit on the neutrino flux from each observed UHECR direction (assuming an equal flux from all of them 
and for $E^{-2}$ energy spectrum) is 
4.99$\times$10${^{-8}}$GeVcm$^{-2}$s$^{-1}$.





\section{\large Acknowledgments}
The authors acknowledge the financial support of the funding agencies:
Centre National de la Recherche Scientifique (CNRS), Commissariat
\'a l'\'ene\-gie atomique et aux \'energies alternatives (CEA), Agence
National de la Recherche (ANR), Commission Europ\'eenne (FEDER fund 
and Marie Curie Program), R\'egion Alsace (contrat CPER), R\'egion 
Provence - Alpes - C\^ote  
d'Azur, D\'e\-par\-tement du Var and Ville de 
La Seyne - sur - Mer, France; Bundesministerium f\"ur 
Bildung und Forschung 
(BMBF), Germany; Istituto Nazionale di Fisica Nucleare (INFN), Italy; 
Stichting voor Fundamenteel Onderzoek der Materie (FOM), Nederlandse 
organisatie voor Wetenschappelijk Onderzoek (NWO), the Netherlands; 
Council of the President of the Russian Federation for young scientists 
and leading scientific schools supporting grants, Russia; National 
Authority for Scientific Research (ANCS - UEFISCDI), Romania; Ministerio 
de Ciencia e Innovaci\'on (MICINN), Prometeo of Generalitat Valenciana 
and MultiDark, Spain. We also acknowledge the technical support of 
Ifremer, AIM and Foselev Marine for the sea operation and the CC-IN2P3 
for the computing facilities.
The authors acknowledge comments and suggestions of members of the Pierre Auger
collaboration.

\label{}





\bibliographystyle{aa}

\bibliography{lela}


\end{document}